# 10 Inventions on modular keyboards
## A TRIZ based analysis

**Umakant Mishra**

Bangalore, India

umakant@trizsite.tk

http://umakant.trizsite.tk

**Contents**

# 1. Introduction

A keyboard is the most important input device for a standard computer. But today's keyboard is not merely a keyboard rather performs a various different activities which were never thought at the time when a keyboard was first invented.

As a standard keyboard is quite spacious many inventions try to use the space of keyboard to use for various activities. The objective of the inventions is to make a keyboard modular, so that specific parts of the keyboard can be detached or attached as per the need.

The advantages of a modular keyboard are as follows:

- Easy to attach and detach external components.
- Break the keyboard and remove the part that is not required, thus giving an advantage of space.
- Expanding the parts of the keyboard thereby giving the advantage of size.
- Make the keyboard folding thereby making it easy to carry
- Attach mouse, telephone, speakers etc. to the keyboard.

## 2. Inventions on modular keyboard

### 2.1 Modular keyboard (Patent 5865546)

**Background problem**

The modern keyboards are having more number of special purpose keys. The 83 keys of the early days' keyboards have been increased to 101 keys to include additional control and function keys. But this enhancement of the keyboard has certain drawbacks. (i) There are two sets of navigation keys out of which most users use only either of the sets. (ii) Besides the user has to use one or more external pointing devices each having a separate cable to connect to the rear side of the computer making the table messy. (iii) Moreover each device needs additional desk space.

**Solution provided by the invention**

Ganthier et al. found the solutions to all the above by disclosing a modular keyboard (US patent 5865546, Assigned to Compaq, Issued in Feb 99). The modular keyboard has several openings in which any of the input modules can be inserted. The user can insert a module and replace a module independently.

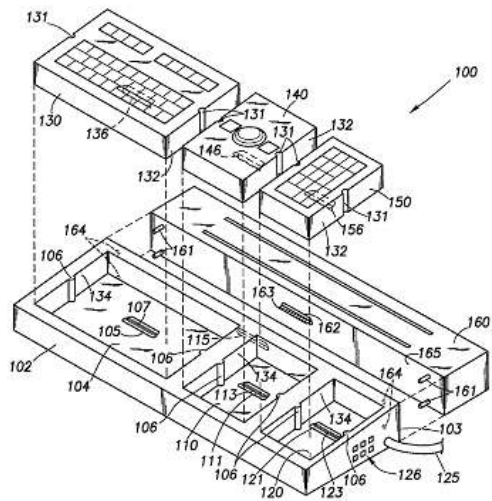

The disclosed modular keyboard has two additional slots other than the basic keyboard module. The user can use any two extra devices depending on need, either a numeric pad and a trackball, or a trackball and joystick or any two other devices depending on need. The user can plug in and plug out any device as required. A controller in the keyboard assembly will determine which types of input device modules are coupled with the keyboard assembly. This has the advantage of no extra cables, quick reconfiguration and saving of space.

**TRIZ based analysis**

Ideally the same keyboard should provide functions of a pointing device, navigation keys, joystick and other devices without needing extra cables and desk space for each device required **(Ideal Final Result)**.

The invention uses the same keyboard cable for all other external devices thus avoiding separate connections for each device to the computer **(Principle-6: Universality)**.

The invention proposes to connect up to two attachments at a time and remove the other attachments that are not used **(Principle-2: Taking out)**.

The invention allows connecting multiple pointing devices of our choice into some sockets on the keyboard **(Principle-7: Nesting)**.

The user can select any two devices of our choice out of various devices including trackball, touchpad, joystick, pointing stick, or stylus etc. to plug into the keyboard **(Principle-15: Dynamize)**.

**2.2 Elevated separate external keyboard apparatus for use with portable computer (Patent 5894406)**

**Background problem**

The keyboards of the portable computers are relatively smaller than the conventional keyboards, which is not convenient for typing for long time. To overcome this problem, some laptops have keyboard ports to attach a regular keyboard for typing comfort. However this solution uses extra desk space and leads to visual and operational limitations.

**Solution provided by the invention**

Michael Blend and Allan Lichtenberg (Patent 5894406, April 99) disclosed a method of embedding a larger external keyboard on the laptop. The external keyboard is specially built to have a cavity to fit on key plate of the portable computer. In other words the external keyboard is kept on the laptop keyboard resting on it. This solves the problem of occupying extra space for the external keyboard, and solves the visual and operational problems of using an external large keyboard.

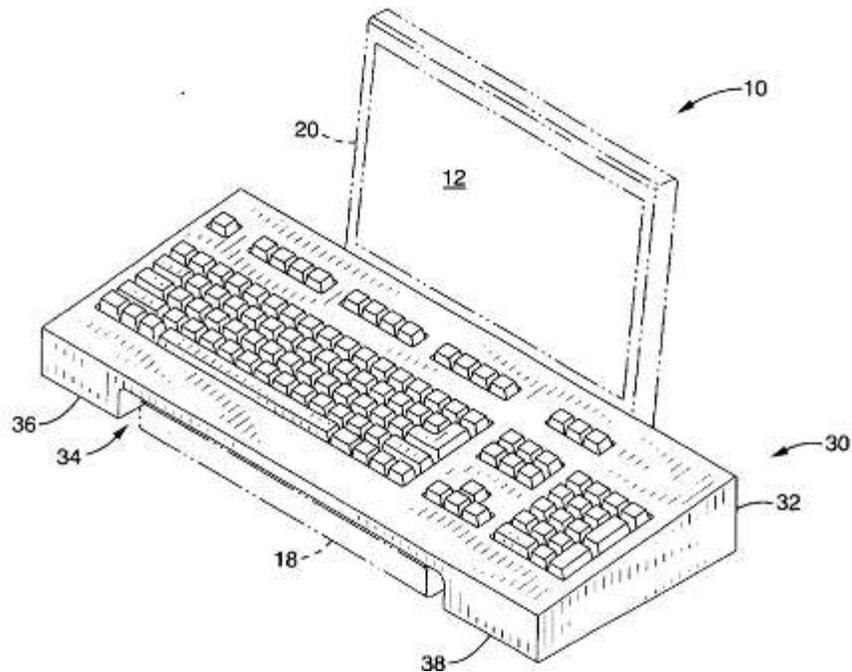

**TRIZ based analysis**

We need a large external keyboard to operate with a portable computer but we don't want to waste the space for two keyboards **(Contradiction)**. We want to use an external keyboard but we don't want to have excessive space between the screen and the keyboard which would cause difficulties in viewing both the monitor and keyboard. **(Contradiction)**.

The invention discloses a keyboard which sits on the laptop keyboard without wasting additional desk space **(Principle-7: Nesting)**.

**2.3 Configurable keyboard to personal computer video game controller adapter (Patent 5896125)**

**Background problem**

Typically joysticks are used to control the video games on the computers. If there are two players to play a game, it needs two such adapters to take input from individual players. Typically only one joystick can be connected to a game port. Games needing multiple game ports create problem.

**Solution provided by the invention**

Niedzwiecki invented an adapter (Patent 5896125, April 99) which has got a socket for the keyboard and sockets for multiple joysticks. The system will contain a user defined association list to map the keys of the game port that of the keyboard. During the time of operation, the input from the game port is sent to the keyboard port based on the association list.

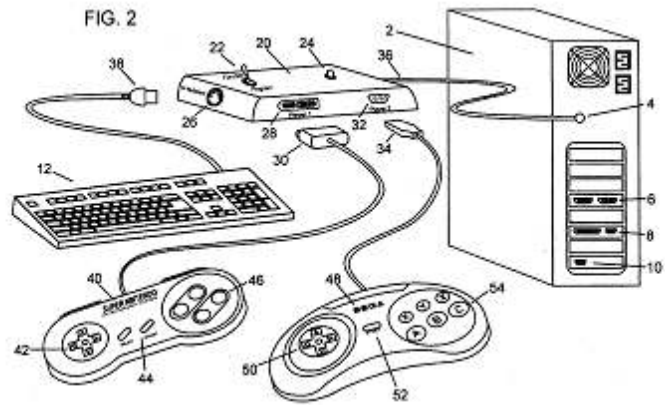

FIG. 2

**TRIZ based analysis**

Ideally the computer should support multiple joysticks for multiple players **(desired result)**.

The invention attaches multiple joysticks to the keyboard **(Principle-5: Merging)**.

The joystick control signals are converted to keyboard control signals **(Principle-28: Mechanics Substitution)** for computer input.

**2.4 Tri-fold personal computer with touchpad and keyboard (Patent 5926364)**

**Background**

The notebook computers are small, light but expensive. The desktop computers are powerful, cheaper but large. There is a need to combine the elements of both desktop and a notebook computer and get the benefits of both.

**Solution provided by the invention**

Karidis disclosed a solution (patent 5926364, assigned to IBM, July 99) of a hybrid packaging design for a portable personal computer. The invention comprises a tri-fold mechanical structure with a touch-screen display screen and a detachable keyboard, which is easy to pack with the computer. It has a stylus, a wireless remote control and other devices as well. The user can operate through the touchpad using fingers or a stylus.

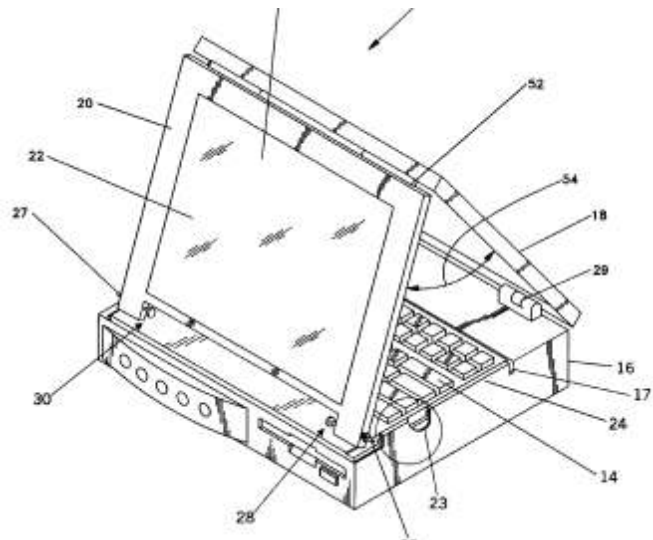

**TRIZ based analysis**

The small notebook computer should contain all the features of a desktop computer **(desired result)**.

Normally a notebook has a folded display unit **(Principle-15: Dynamize)**. The invention proposes to increase the number of folds to threefold **(Principle-17: Another dimension)**.

It uses a detachable keyboard **(Principle-15: Dynamize)**.

**2.5 Portable computer having split keyboard and screen (Patent 5949643)**

**Background problem**

The classical problem exists as usual with the size of the keyboard for a portable computer. The small computing devices like PDA and notebook do not have space to accommodate large keyboards. But small keyboards are neither efficient nor comfortable for typing. Although there are several inventions attempting to solve the same problem, it is still required to find a better solution.

**Solution provided by the invention**

Batio invented a folding keyboard (US patent 5949643, Sep 99) for the PDA, notebook and other small computing devices. The keyboard consists of two, pivotally hinged halves, which make a perfectly flat horizontal keyboard. Each half has its own set of keys and space bar. The keyboard also has a pointing device and a joystick adapter. The invention also provides a dual split screen where each half of the split screen is pivotally mounted for universal rotation.

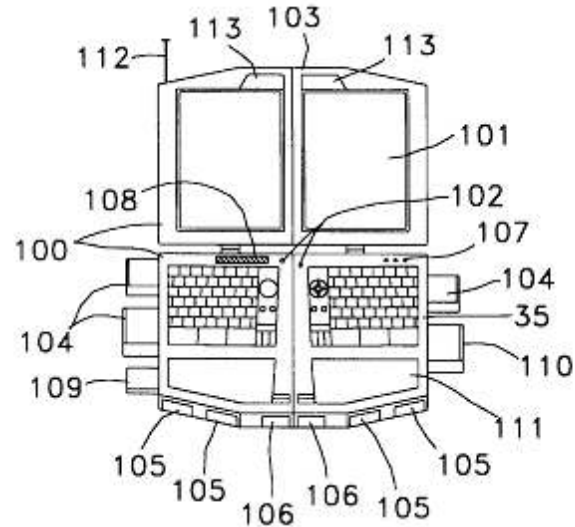

This invention is not confined to only the folding nature of the keyboard. It also has two independent dedicated microprocessors. While one microprocessor use used as a notebook computer, the other microprocessor may be dedicated for video games or a set-top converter box for Internet access through cable TV.

**TRIZ based analysis**

The invention splits the keyboard into two halves. It also splits the screen into two halves **(Principle-1: Segmentation)**.

The two halves of the keyboard and screen have folding joints, which make it possible to open during usage and fold during storage **(Principle-15: Dynamize)**.

**2.6 Portable computer with ergonomic keyboard (Patent 6304431)**

**Background problem**

Conventionally the keyboard is statically attached to a portable computer. This needs the user to position himself straight to operate the computer which may not be hygienic or comfortable. There are some ergonomic keyboards for the desktops but nothing for the laptops. There is a need for an ergonomic keyboard for laptop.

**Solutions provided by the invention**

Myung-Jung Kim invented an ergonomic keyboard (Patent Number: 6304431, assigned to Samsung Electronics, Oct 01). The invention makes the keyboard more ergonomic without surrendering the portability.

According to the invention, the keyboard unit is attached to an upper surface of the main body and is structurally and operationally separated along a central axis into a left keyboard half and a right keyboard half. Both the keyboard halves are coupled to the main body but rotate on the first and second hinges to be suitably positioned according to user requirement.

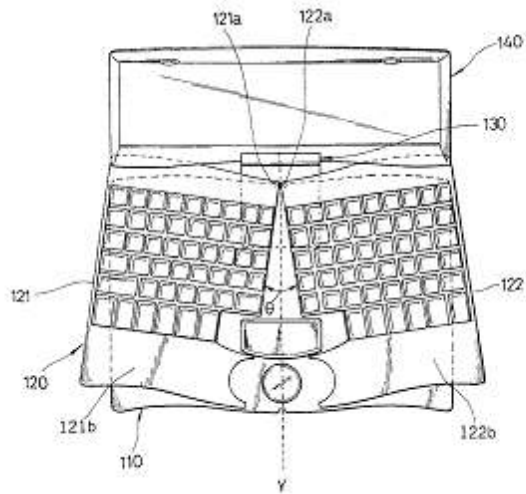

**TRIZ based analysis**

Solution: Split the keyboard into two, one for each arm **(Principle-1: Segmentation)**, and make them flexible to adjust with the angle of the arm. **(Principle-15: Dynamize)**.

**2.7 Detachable keyboard (Patent 6317061)**

**Background problem**

Different computer users choose different types of keyboard, some want a simple keyboard with alphanumeric keys, and others want a lot of additional functions. Sometimes a partial keyboard is convenient more beneficial, for example the game players may need only cursor movement keys.

**Solution provided by the invention**

Batra et al. invented a keyboard (Patent 6317061, Nov 2001) invents a keyboard that allows to switch between a full size keyboard and a partial keyboard. The keyboard comprises of a first keyboard having first set of functions, and a docking station having second set of functions. The first keyboard is coupled to the keyboard docking station to make a complete keyboard. The partial keyboard allows the user to remove it from the detachable keyboard and operate it as an input device but at reduced functions. This permits the user to choose between a complete keyboard and a partial keyboard.

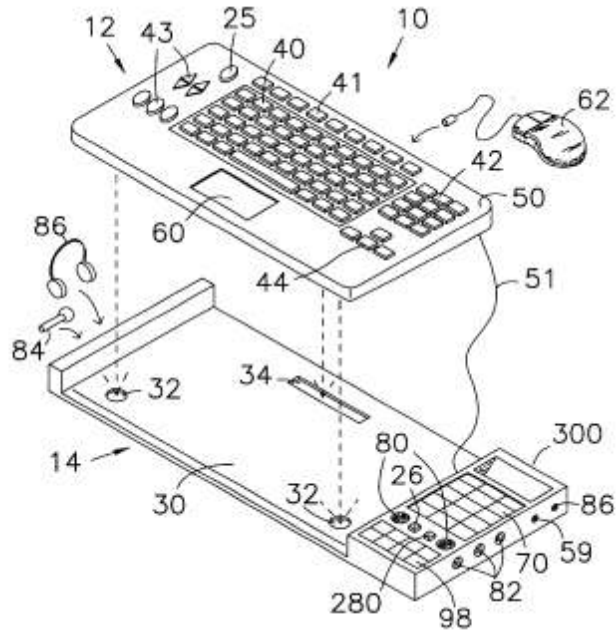

**TRIZ based analysis**

The keyboard should have all the features when the user requires them, and they keyboard should have only partial features when other features are not required **(Contradiction)**.

The invention makes two attachable units **(Principle-1: Segmentation)**.

The detachable extension can be attached to make a complete full featured keyboard **(Principle-15: Dynamize)**.

**2.8 Keyboard with multiple input/output function (Patent 6542092)**

**Background problem**

There is a growing trend to access the Internet by using a set-top-box equipped with a wireless keyboard. There is also a growing trend to use mobile handsets to communicate with family members and other individuals. It is desirable to have a keyboard which can interface with the set-top-box as well as support the mobile handset.

**Solution provided by the invention**

Jung-Chuan Pan invented a keyboard (patent 6542092, assigned to Silitek Corporation, April 2003) having multiple input/output functions. The keyboard can be connected to a computer (or a set-top-box).
Besides, the keyboard also contains a handset slot and integrates the wireless communication function of the mobile phone.

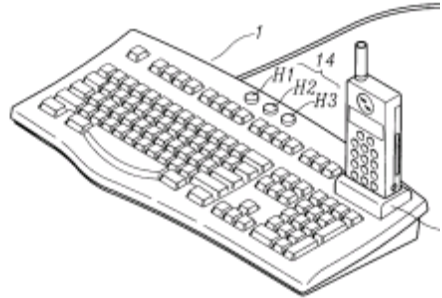

### TRIZ based analysis

The invention provides various communication interfaces to the keyboard. It provides a socket for holding and charging the mobile phone **(Principle-31: Hole, Principle-38: Enrich)**.

### 2.9 Keyboard with detachably attached unit having multimedia key function (Patent 6545668)

**Background problem**

The multimedia keyboards are typically larger as they contain a few special keys for different kinds of multimedia functions. As the keys are primarily built into the keyboard the keyboard remains larger even if the multimedia functionalities are not used. We need a mechanism to make the keyboard smaller when the multimedia functionalities are not used.

**Solution provided by the invention**

Hayama invented a keyboard (patent 6545668, assigned to Fujitsu Takamisawa, April 2003) which contains an auxiliary keyboard for multimedia functions which is detachable from the main keyboard. When the multimedia key functions are not utilized the auxiliary keyboard unit can be detached from the main keyboard unit to save space.

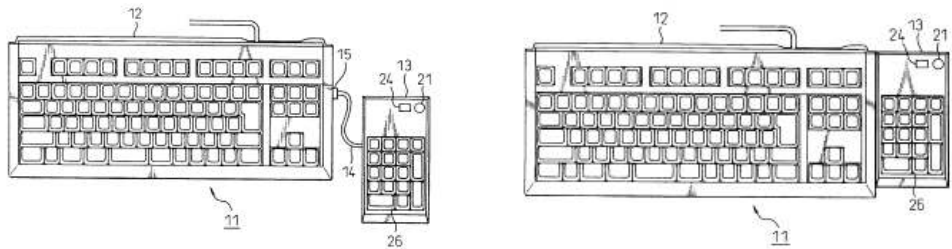

According to the invention the auxiliary keyboard comprises of ten keys for different multimedia functions. The multimedia function definitions are stored in a table in the main computer, which is read by the CPU when powered on.

**TRIZ based analsysis**

The same keyboard should provide multimedia functionalities without needing additional space **(desired result)**.

The invention includes multimedia functions in a small keyboard and uses the small keyboard as an attachable unit to the main keyboard **(Principle-15: Dynamize)**.

## 2.10 Snap-on keyboard and method of integrating keyboard (Patent 6573843)

### Background problem

Desktop computers are designed with detachable input/output devices such as monitors, keyboards, mice etc. that allows it to configure in the most desirable position and most ergonomic manner. In contrast, most components of the laptops are fixed. Some portable computers provide an auxiliary keyboard port to connect an external keyboard but an external keyboard covers a lot of desk space. There is a need to provide an ergonomic keyboard to portable computers without affecting its portability.

### Solution provided by the invention

Stephen Murphy invented a detachable snap-on keyboard (Patent 6573843, Assigned to Micron Technology, June 03) for the laptops. The snap-on keyboard can be placed over the built in keyboard without needing additional table space.

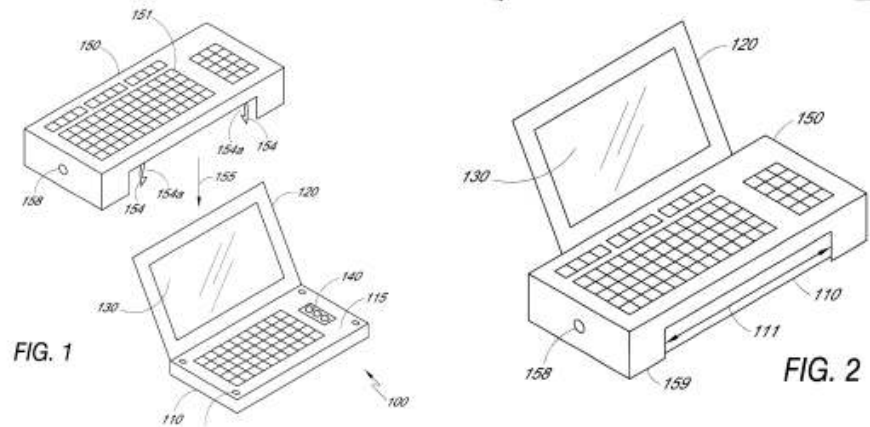

The snap-on keyboard is attached to a connector on the laptop. The connector is configured to detect the snap-on keyboard when it is kept on the surface of the laptop and pressed downward.

### TRIZ based analysis

The invention uses a snap-on keyboard, which is placed on the laptop keyboard **(Principle-7: Nested Doll)**.

The laptop keyboard has a cavity or slot on the top to connect the external snap-on keyboard **(Principle-31: Hole)**.